# Metasurface Smart Glass for Object Recognition

Cheng-Chia Tsai*[1], Xiaoyan Huang*[1], Zhicheng Wu*[2], Zongfu Yu[2] and Nanfang Yu[1]

[1]Department of Applied Physics and Applied Mathematics, Columbia University, New York, NY 10027, USA.

[2]Department of Electrical and Computer Engineering, University of Wisconsin-Madison, Madison, WI 53706, USA

*These authors contribute equally to the work.

**Abstract**

Recent years have seen a considerable surge of research on developing heuristic approaches to realize analog computing using physical waves. Among these, neuromorphic computing using light waves is envisioned to feature performance metrics such as computational speed and energy efficiency exceeding those of conventional digital techniques by many orders of magnitude. Yet, neuromorphic computing based on photonics remains a challenge due to the difficulty of training and manufacturing sophisticated photonic structures to support neural networks with adequate expressive power. Here, we realize a diffractive optical neural network (ONN) based on metasurfaces that can recognize objects by directly processing light waves scattered from the objects. Metasurfaces composed of a two-dimensional array of millions of meta-units can realize precise control of optical wavefront with subwavelength resolution; thus, when used as constitutive layers of an ONN, they can provide exceptionally high expressive power. We experimentally demonstrate ONNs based on single-layered metasurfaces that modulate the phase and polarization over optical wavefront for recognizing optically-coherent binary objects, including hand-written digits and English alphabetic letters. We further demonstrate, in simulation, ONNs based on metasurface doublets for human facial verification. The advantageous traits of metasurface-based ONNs, including ultra-compact form factors, zero power consumption, ultra-fast and parallel data processing capabilities, and physics-guaranteed data security, make them suitable as "edge" perception devices that can transform the future of image collection and analysis.

**Introduction**

Object recognition is being exploited in an increasingly wide range of applications, such as image annotation, vehicle counting and tracking, pedestrian detection, and facial detection and recognition. Using digital images from cameras and videos and machine learning models, computer vision recognizes objects by translating high-dimensional visual signals from the real world into lower dimensional representations. The full technology stack in this conventional approach requires a compound optical system to form images, an optoelectronic sensor for analog-to-digital conversion, and digital processors to implement artificial neural networks (ANNs) [1-3]. Consequently, the resulting system is bulky and power hungry, reacts slowly due to the latency between technology modules, and is vulnerable to cyber-attack. These problems are exacerbated as the demand for high power efficiency, computational speed, and data security increases rapidly with the explosion of data volume and the wide availability of mobile devices with computer vision features.

An emerging physical platform for object recognition is an optical neural network (ONN). An ONN utilizes photonic elements and circuits to form a layered architecture emulating that of digital ANNs to directly process optical signals from target objects [4-21]. In an ideal ONN, optical signals are manipulated by layers of elements in sequence that perform linear transformations and nonlinear activations, which are pre-trained to enable the network to perform device-specific computing tasks. ONNs have a computational speed characterized by the propagation speed of light; it can be entirely passive, requiring no additional power after the optical input is generated. Furthermore, computational algorithms are hard coded in the intrinsic and engineered materials of ONNs so that data security is guaranteed.

ONNs have been demonstrated in a number of optical platforms (**Table S1**) [4-21]. Following the traditional layered architecture of digital ANNs, ONNs based on integrated photonic circuits have been realized, where signals are encoded in an array of pulses and linear transformation is conducted by a network of tunable Mach-Zehnder interferometers (MZIs) [4-9]. In this approach, the large size of MZIs compared to optical wavelength limits the expressive power and thus the capabilities of the ONN to accomplish complex tasks. Recently, a diffractive ONN based on metamaterials that performs artificial neural computing utilizing the physics of wave dynamics has been proposed [10]. In this approach, signals are encoded in a continuous optical wavefront and the three-dimensional (3D) architecture and materials composition of the metamaterial are trained to recognize objects by scattering light into target zones on an output layer. In this approach, linear transformation and nonlinear activation are realized, respectively, by light scattering at optically linear and nonlinear inclusions of the metamaterial. While in principle 3D metamaterial ONNs can provide a large expressive power, current device manufacturing techniques only allow for reliable fabrication of a two-dimensional (2D) version without nonlinear inclusions, which would offer a comparatively much smaller expressive power.

A more viable approach to realize diffractive ONNs is to condense the depth of the network into a few discrete layers of diffractive components [11-17]. The wavefront exiting a layer can be considered as a collection of point sources, in accordance with the Huygens-Fresnel principle, and forms a diffraction pattern as it propagates to the subsequent layer. The 2D distribution of scattering elements on a diffractive component can be trained to shape a wavefront in phase, amplitude, and polarization by both local modulation and nonlocal interference. This discrete-layered architecture enabling 2D matrix manipulation is ideal for image-based computer vision applications, including object recognition. Previous work has demonstrated such diffractive ONNs, which are composed of multiple layers of "diffractive surfaces" and can classify objects such as hand-written digits and fashion objects [11-15]. The diffractive layers are based on 3D-printed masks operating in the THz spectral range [11-14] or a combination of a digital micromirror device (DMD) and a spatial light modulator (SLM) operating in the visible spectral range [15]. Therefore, the pixels of the diffractive layers cannot have subwavelength sizes, while simultaneously modulating all properties of light (phase, amplitude, and polarization), which limit the expressive power of the ONNs. Furthermore, the utility of the neural networks would be hampered by their large dimensions and a lack of wide availability of SLMs and THz sources and detectors. More recently, hybrid optoelectronic neural networks that combine the strength of an optical frontend in linear transformation and that of a digital neural network in nonlinear activation have been demonstrated [18-21]. However, the hybrid neural networks still have power consumption and latency issues, which are inherent to their reliance on conducting neural computing partially with electronics.

Here, we propose and demonstrate a diffractive ONN based on metasurfaces, dubbed a metasurface "smart glass", that directly processes light waves scattered by an object using its internal nanostructures. A metasurface [22-27] is a 2D version of a metamaterial that utilizes strong interactions between light and 2D nanostructured thin films to control light in desired ways, realizing device functions such as flat lenses and holograms [25-27]. Metasurfaces are typically composed of a 2D array of nano-pillars (i.e., "meta-units") of various cross-sectional shapes and can offer complete and precise manipulation of optical phase, amplitude, and polarization across the wavefront with sub-wavelength resolution. The collective response of millions of sub-wavelength meta-units enables efficient parallel computing with a high level of expressive power; as a result, tasks typically solved using a complex, multi-layered network can be accomplished by our smart glass using a metasurface singlet or doublet. The metasurfaces are manufactured by CMOS-compatible nanofabrication techniques and can enable miniaturization of the discrete-layered diffractive neural networks operating in the optical spectral range, where the light sources and detectors are readily available. Our metasurface smart glasses do not need any power supply or digital processor: they act as passive computing devices that operate at the speed of light. The potential of metasurface-based neural networks has been recently demonstrated in classification of simple binary objects such as digits and fashion objects [28].

In this work, we investigate the computational capacity of metasurface-based diffractive networks by experimentally demonstrating smart glasses for a few recognition tasks using single-layered metasurfaces that modulate the phase and polarization of the optical wavefront. We realize recognition of four classes of hand-written digits with an accuracy exceeding 99% and recognition of ten classes of hand-written digits with an accuracy of approximately 80%. We further implement single-layered polarization-multiplexing smart glasses to solve more complex tasks, for example, recognizing alphabetical letters using light at one polarization state and their typographic styles (i.e., normal or italic) using light at the orthogonal polarization state with accuracies exceeding 90%. Lastly, we theoretically investigate the capability of metasurface smart glass doublets in performing advanced recognition tasks and demonstrate human facial verification with an accuracy of approximately 80%, which is comparable to that achieved by a conventional digital ANN with three convolutional layers.

**Results**

**Training and experimental implementation of single-layered metasurface smart glass**

**Figure 1a** illustrates the working principle of the metasurface smart glass using a specifical example. An input object, a hand-written digit "4", upon excitation of an incident coherent light beam, generates an optical wavefront with characteristic amplitude and phase profiles (**Figs. 1b and 1c**). This complex optical wavefront, propagating over a certain distance (i.e., object distance), is then processed by a metasurface, which superimposes a phase modulation to the wavefront (**Fig. 1d**). The modulated light wave further propagates over a certain distance (i.e., imaging distance) in the forward direction and produces an optical diffraction pattern that lights up a few predefined zones on the detection plane. The zone that receives the highest optical intensity, in this particular example, identifies the initial object. The input object, the metasurface, and the detection plane represent, respectively, an input layer, a hidden layer, and an output layer of a neural network, and every pixel in either one of the three layers represent an artificial neuron. In this configuration, each neuron in the hidden layer is connected to all the neurons in the input layer via optical interference, and each neuron in the output layer is similarly connected to all the neurons in the hidden layer. The optical interference provides a form of nonlinear activation by generating cross-products of optical wavelets. The phase modulation at each neuron of the hidden layer represents a trainable linear transformation. Object recognition is accomplished by training all the neurons in the hidden layer to maximize the light intensity within a specific zone of the output layer, depending on the classification label of the input object.

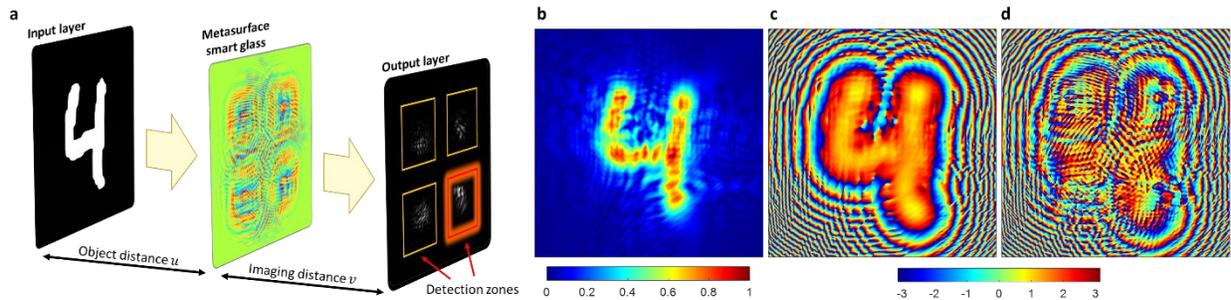

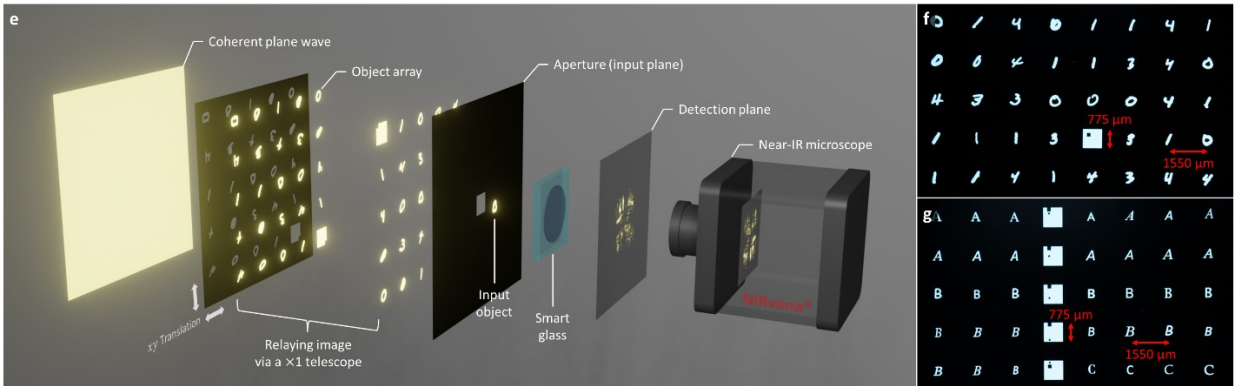

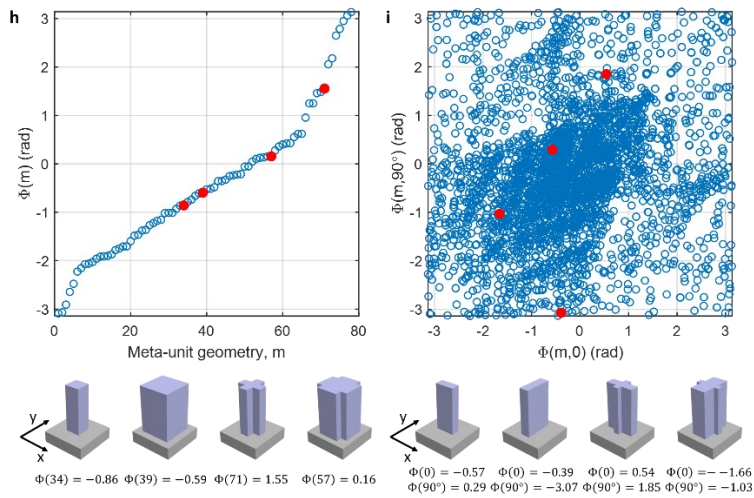

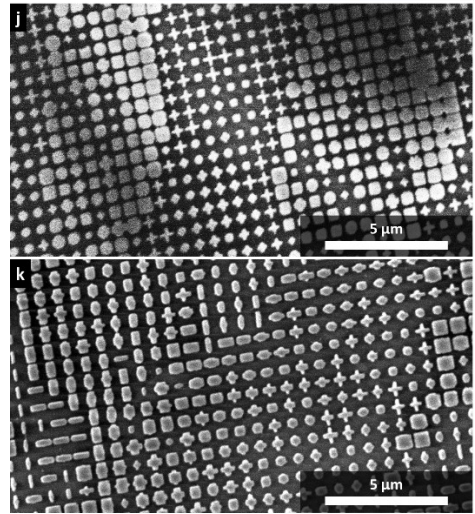

**Figure 1 | Operation of a metasurface smart glass for object recognition.** (a) Schematic illustrating that a smart glass channels light from an input image preferentially onto one of several detection zones on the output plane. The yellow arrows indicate the direction of light propagation. The squares on the output plane define the detection zones where the zone (highlighted in red) corresponding to the identity of the input object receives the highest share of intensity. (b) and (c) Amplitude and phase profiles over the optical wavefront generated by the object (a hand-written "4") before it enters the metasurface smart glass. (d) Modulated phase distribution over the optical wavefront as it exits the metasurface. (e) Schematic of the experimental setup. (f) and (g) Microscopic photos of plates containing input objects (hand-written digits and typed letters) and alignment marks. The plates are composed of apertures defined on an opaque photomask. An array of objects is reimaged on the input plane by a ×1 telescope and filtered by an aperture to allow light from a single object to propagate to the metasurface, as shown in (e). (h) Phase responses of meta-units in the isotropic library, where $m$ denotes the index of the meta-units. (i) Phase response of meta-units in the birefringent library. Each blue circle represents one meta-unit. Red solid circles represent a few exemplary meta-units illustrated below. (j) and (k) SEM images of fabricated metasurfaces consisting of isotropic and birefringent meta-units, respectively.

Our ONN is designed for the near-infrared light at $\lambda=1,550$ nm. The input object and the metasurface smart glass both have a dimension of $500\,\lambda \times 500\,\lambda$ and are digitized into $1000 \times 1000$ pixels. The object and imaging distances are both $2000\,\lambda$. The smart glass is composed of a single metasurface modelled as a phase mask with zero thickness on a substrate with a thickness of $322.58\,\lambda$ (~500 μm) and a refractive index of 1.44 (silicon dioxide). During the training process, optically coherent, binary images (e.g., hand-written digits and alphabetic letters) are fed into the neural network and propagation of light waves through the diffractive network is numerically computed using the Rayleigh-Sommerfeld diffraction theory [29,30]. A loss function is defined to evaluate the cross-entropy between the calculated intensity distribution over the detection plane and the target intensity distribution, which is 1 for the zone that matches with the label of the input and 0 elsewhere. The phase profile of the metasurface is iteratively adjusted using a large number of input objects during the training process, where the loss function is minimized using the "Adam" optimization algorithm adapted from the stochastic gradient-based optimization method [31].

Several measures are taken to improve the robustness of the ONN against experimental errors (**Supplementary Section S3**). For example, non-uniform optical illumination to the input objects, random mispositioning of the input object, smart glass, and detection zones, and random variations of the object and imaging distances are included in the training process; an auxiliary term, proportional to the ratio between the intensity in the predefined zones of the detection plane and the total intensity in the detection plane, is subtracted from the overall loss function to increase the contrast of the zones of interest over the optical background.

A schematic of the experimental setup is shown in **Fig. 1e**. A telecom laser beam ($\lambda = 1,550$ nm) is incident on a photomask to create input optical objects. The photomask is made of a black emulsion photo-plotted on a mylar sheet, containing a 2D array of objects (i.e., numerical digits or alphabetic letters) that are transparent within an object and opaque outside of it (**Figs. 1f** and **1g**). The incident beam has a diameter of approximately 3 mm, which is much larger than the size of individual input objects (0.775 mm × 0.775 mm), to minimize non-uniformity in illumination. A motorized translation stage each time moves one input object to the central axis of the optical setup. The input object is relayed by a telescope with a unity magnification and the relayed object is superimposed onto a square aperture (0.775 mm × 0.775 mm), so that stray light from adjacent input objects on the photomask is blocked. The diffraction pattern of the object is then processed by the metasurface smart glass, and the output image is collected by a microscope with an objective focused on the detection plane and measured by an InGaAs camera. The optical intensities in the predefined detection zones are extracted from the image and the identity of the input object is predicted according to the zone receiving the highest intensity.

The metasurface is made of amorphous silicon for its low extinction coefficient in the near-infrared and is composed of meta-units 1 μm in height and arranged in a square lattice with a periodicity of 750 nm on a silicon dioxide substrate. The phase responses of two meta-unit libraries used in this work are shown in **Figs. 1h** and **1i**. The meta-units in one library have a cross-section with four-fold symmetry and are thus optically isotropic; those in the other library have a cross-section with two-fold symmetry and can introduce form birefringence and thus distinct phase responses for incident light with orthogonal polarization states. To test polarization-multiplexing smart glasses, a linear polarizer is inserted in front of the photomask and an object is tested twice with incident light at two orthogonal polarization states. Experimental characterization of the metasurface smart glasses is conducted with 10-40 distinct objects for each classification label to estimate the accuracy of object recognition. The tested objects are haphazardly chosen from a dataset excluded from the dataset used in the training.

**Smart glasses for recognition of hand-written digits**

The first functionality we demonstrate experimentally is to recognize 4 classes of numerical digits, {0, 1, 3, 4}, from the MNIST hand-written digit database. The phase modulation (**Fig. 2a**) is trained to concentrate light scattered from the binary image of a digit into one of the four square zones on the detection plane as defined in **Fig. 2b**. The trained phase modulation is implemented by a metasurface (**Fig. 2c**) based on meta-units that are optically isotropic (**Fig. 2d**); correspondingly, the polarization state of the incident laser beam is not controlled. **Figures 2e-g** show a few exemplary classification cases, where the metasurface smart glass successfully classifies digits on the basis of the resulting intensity distributions on the detection plane. The observed diffraction patterns on the detection plane (**Fig. 2f**) agree well with analytically calculated diffraction patterns (**Fig. 2g**), indicating that the metasurface provides a precise phase modulation consistent with the design. Training of a single-layered digital ANN simulating the architecture of this ONN reports an accuracy of 99.14% (**Fig. 2j**), while measurement of 116 input digits (4 classes and N>25 for each class) results in a recognition accuracy of 99.14% (**Fig. 2k**). The raw data of optical intensity integrated over the four detection zones in the training and testing processes are summarized in the bar charts in **Figs. 2h** and **2i**, showing good agreement between theoretical and experimental results. The zone corresponding to the correct identity of the digit has an integrated intensity approximately 22% higher than that of the other three zones; this large inter-zone intensity difference ensures the robustness of our ONN against experimental variations.

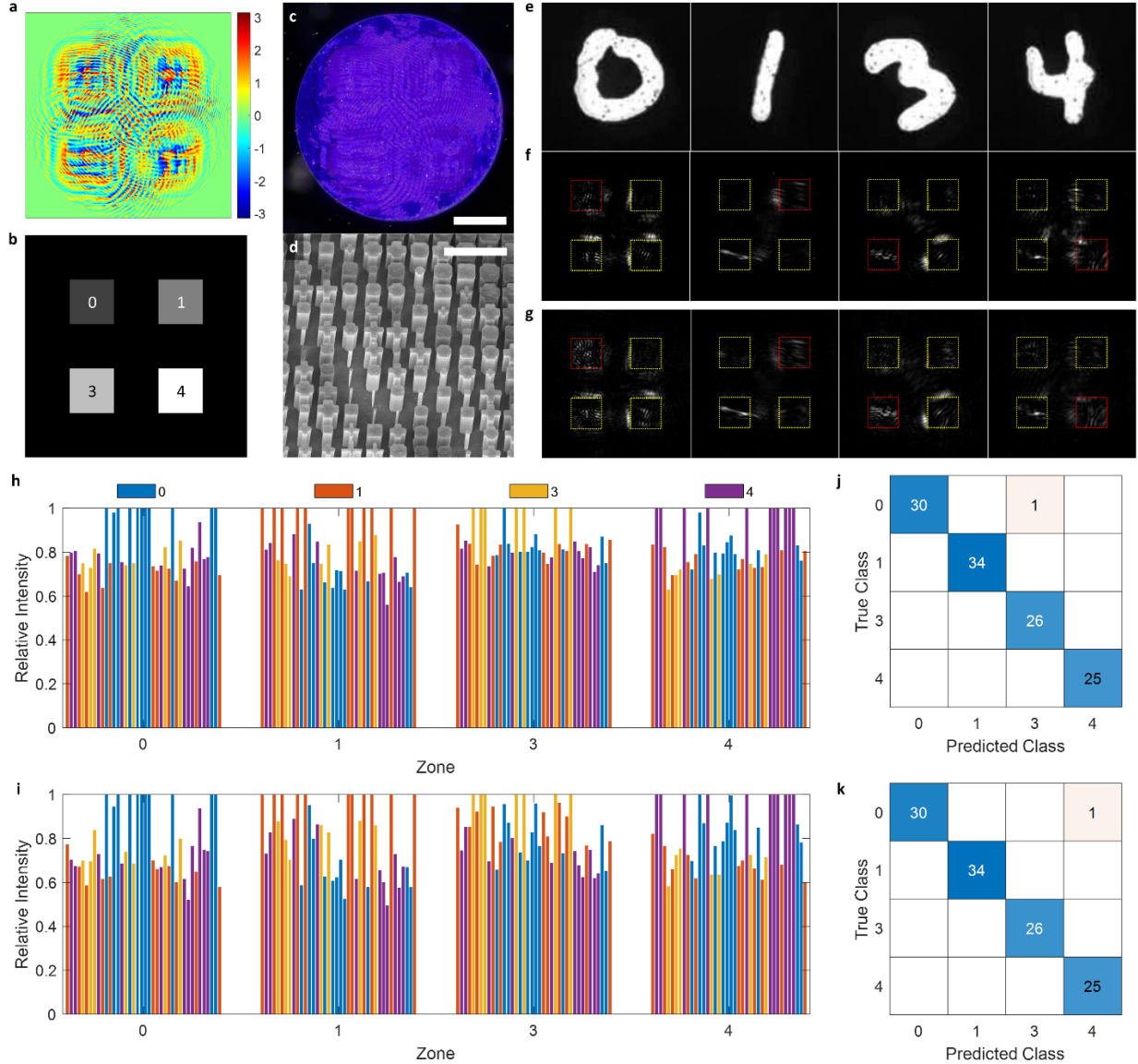

**Figure 2 | Recognition of four classes of hand-written numerical digits.** (a) Trained phase modulation on the metasurface. (b) Detection zones on the output plane. (c) Optical microscopic photo and (d) SEM image of the fabricated metasurface. Scale bars: 200 μm in (c) and 2 μm in (d). (e)-(g) Examples showing the recognition of four hand-written digits, "0", "1", "3", and "4": (e) Input images; (f) Measured intensity distributions on the output plane showing that the detection zone (red) corresponding to the true classification of the digits has the highest integrated optical intensity; (g) Analytically calculated intensity distributions on the output plane showing a high degree of consistency with the measured results. (h) Theoretical integrated intensities of the four detection zones for 40 randomly selected input digits, with the highest value normalized to 1. The colors of bars represent the true classification of the digits. (i) Experimental integrated intensities

of the four detection zones for the same 40 digits as those shown in (h), with the highest value normalized to 1. (j) and (k) Confusion matrices summarizing the theoretical and experimental results of recognizing 116 hand-written "0", "1", "3", and "4", respectively.

We next explore classification of all 10 classes of hand-written digits using a single-layered metasurface smart glass. The trained optical phase profile of the metasurface is shown in **Fig. 3a** and it is implemented similarly using a metasurface based on optically isotropic meta-units (**Fig. 3d**). The 10 circular detection zones are arranged in a circular array on the detection plane (**Fig. 3b**). Three examples of classification are shown in **Figs. 3e-g**. This ONN has an experimental recognition accuracy of 78.37% (**Figs. 3i** and **3k**) based on measurement of 208 input digits (10 classes and N>10 for each class). The experimental accuracy is lower than the theoretical accuracy of 86.50% (**Figs. 3h** and **3j**) according to results of the training process, suggesting that the ONN has a reduced robustness against experimental errors. In fact, the intensity contrast between the detection zones with the highest intensity and the second-highest intensities reduces to ~10% (**Figs. 3h** and **3i**) as the number of zones increases to 10, making the ONN more susceptible to experimental errors such as the non-uniformity of the incident beam, and mispositioning of the components and the detection zones of the ONN.

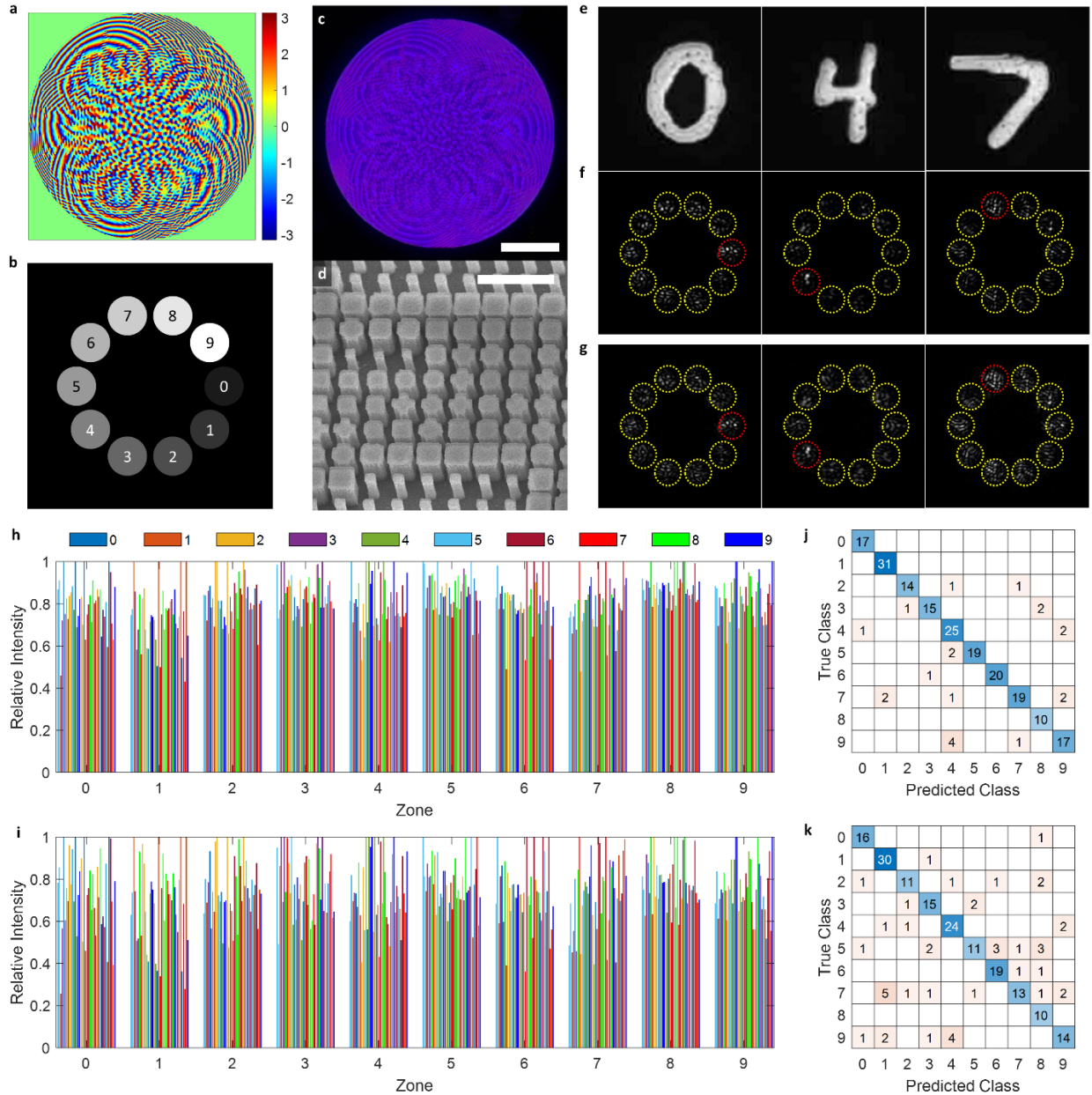

**Figure 3 | Recognition of 10 classes of hand-written numerical digits.** (a) Trained phase modulation on the metasurface. (b) Detection zones on the output plane. (c) Optical microscopic photo and (d) SEM image of the fabricated metasurface. Scale bars: 200 μm in (c) and 2 μm in (d). (e)-(g) Examples showing the recognition of three hand-written digits, "0", "4", and "7": (e) Input images; (f) Measured intensity distributions on the output plane showing that the detection zone (red) corresponding to the true classification of the digits has the highest integrated optical intensity. (g) Analytically calculated intensity distributions on the output plane showing a high degree of consistency with the measured results. (h) Theoretical integrated intensities of the 10

detection zones for 40 randomly selected input digits, with the highest value normalized to 1. The colors of bars represent the true classification of the digits. (i) Experimental integrated intensities of the 10 detection zones for the same 40 digits as those shown in (h). (j) and (k) Confusion matrices summarizing the theoretical and experimental results of recognizing 208 hand-written digits, respectively.

**Polarization multiplexing and multitasking smart glasses**

10-digit recognition is computationally a more expensive task than categorizing only 4 classes of digits. We devise a polarization-multiplexing strategy to reduce the complexity of the task by dividing the 10 digits into two groups and performing the recognition task using light linearly polarized in orthogonal directions: horizontally polarized light for recognizing digits {1, 3, 4, 7, 8} and vertically polarized light for recognizing digits {0, 2, 5, 6, 9}. The smart glass is constructed using the birefringent meta-unit library (**Fig. 1i**) to provide distinct phase modulations for light polarized in orthogonal directions (**Fig. 4a**). Two examples for recognizing digits "4" and "6" are illustrated in **Figs. 4e-g**. The training process reports accuracies attaining 94.80% and 94.00% for the two groups of digits, respectively (**Fig. 4h**); recognition accuracies achieved based on measurement of 111 input digits belonging to the first group and 97 digits belongs to the second group are 90.99% and 81.44%, respectively (**Fig. 4i**), which are substantially higher than that of the non-birefringent device (**Fig. 3k**). We note that the phase coverage provided by the birefringent meta-unit library is more discrete than that of the isotropic meta-unit library; therefore, the phase responses of the fabricated birefringent metasurface deviate from the desired phase profiles more than does the non-birefringent device. This issue can be addressed by including more archetypes of meta-units in the library (only rectangle and cross motifs are used currently). We also note that the function demonstrated in **Fig. 4** is not entirely 10-class classification, because a digit has to be pre-categorized into one of two groups. We find that an ONN composed of a single birefringent metasurface with a dimension of 1000 × 1000 pixels has insufficient expressive power to experimentally accomplish the 10-class classification task with an accuracy higher than 90% (**Supplementary Section S3**).

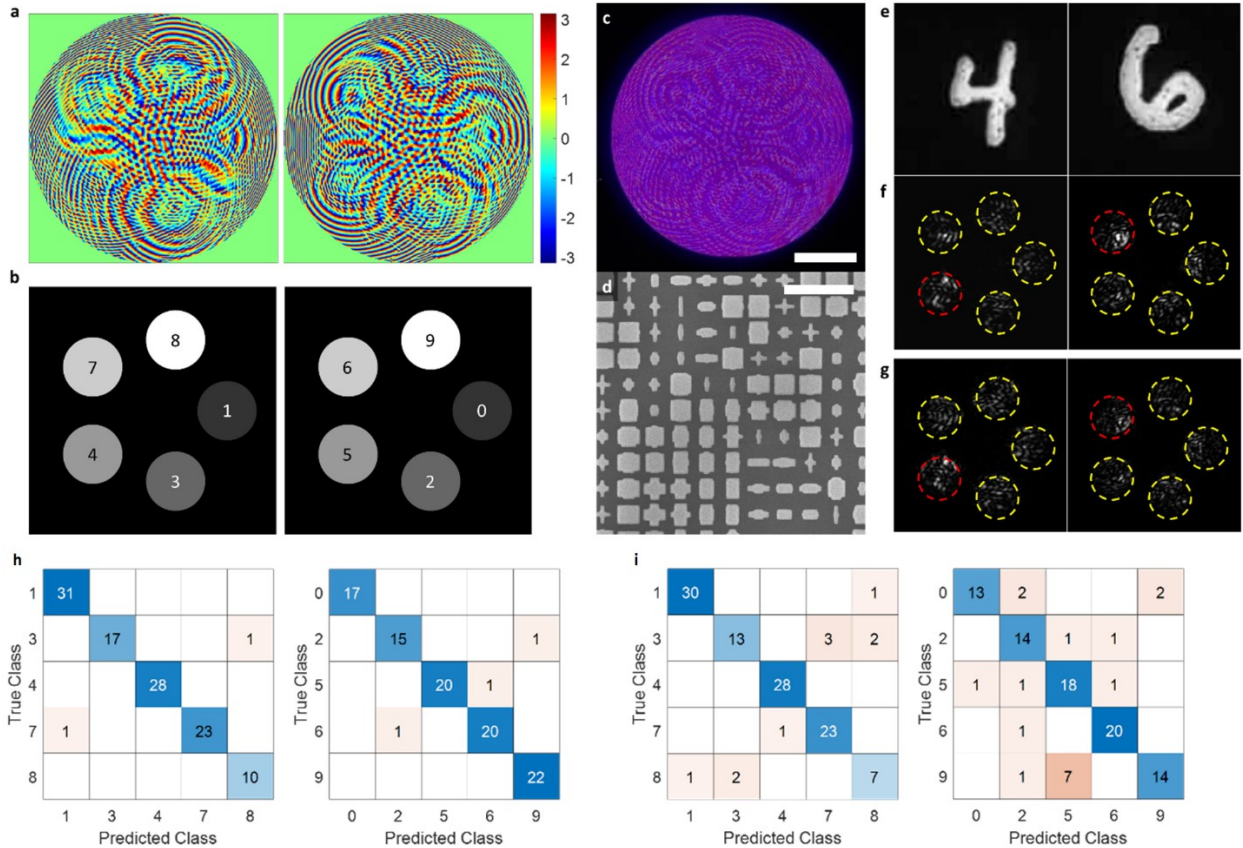

**Figure 4 | Recognition of 10 classes of hand-written numerical digits using a polarization-multiplexing smart glass.** (a) Trained phase modulations on the metasurface and (b) arrangements of detection zones on the output plane at two orthogonal incident polarizations for recognizing, respectively, two groups of digits: {1, 3, 4, 7, 8} and {0, 2, 5, 6, 9}. (c) Optical microscopic photo and (d) SEM image of the fabricated metasurface. Scale bars: 200 μm in (c) and 2 μm in (d). (e)-(g) Two examples showing the recognition of hand-written "4" and "6" using light with orthogonal polarization states: (e) Input images; (f) Measured intensity distributions on the output plane showing that the detection zone (red) corresponding to the true classification of the digits has the highest integrated optical intensity. (g) Analytically calculated intensity distributions on the output

plane showing a high degree of consistency with the measured results. (h) and (i) Confusion matrices at two orthogonal polarization states summarizing the theoretical and experimental results of recognizing 208 hand-written digits, respectively.

We further utilize polarization multiplexing to realize a multi-tasking metasurface smart glass that classifies typed alphabetical letters and simultaneously distinguishes the typographic styles of the letters (**Fig. 5**). Specifically, when incident illumination is polarized in the horizontal direction, scattered light from a letter with a certain font is modulated by the smart glass to preferentially light up one of the four zones on the detection plane, corresponding to 4 letters: {A, B, C, D}; scattered light polarized in the vertical direction, however, falls in one of the two zones in the upper row to indicate if the letter is normal or italic. Experiments using 168 inputs (4 letters each with 21 fonts and 2 typographic styles) demonstrate accuracies of 92.81% and 100% for letter classification and typographic style recognition, respectively (**Fig. 5j**).

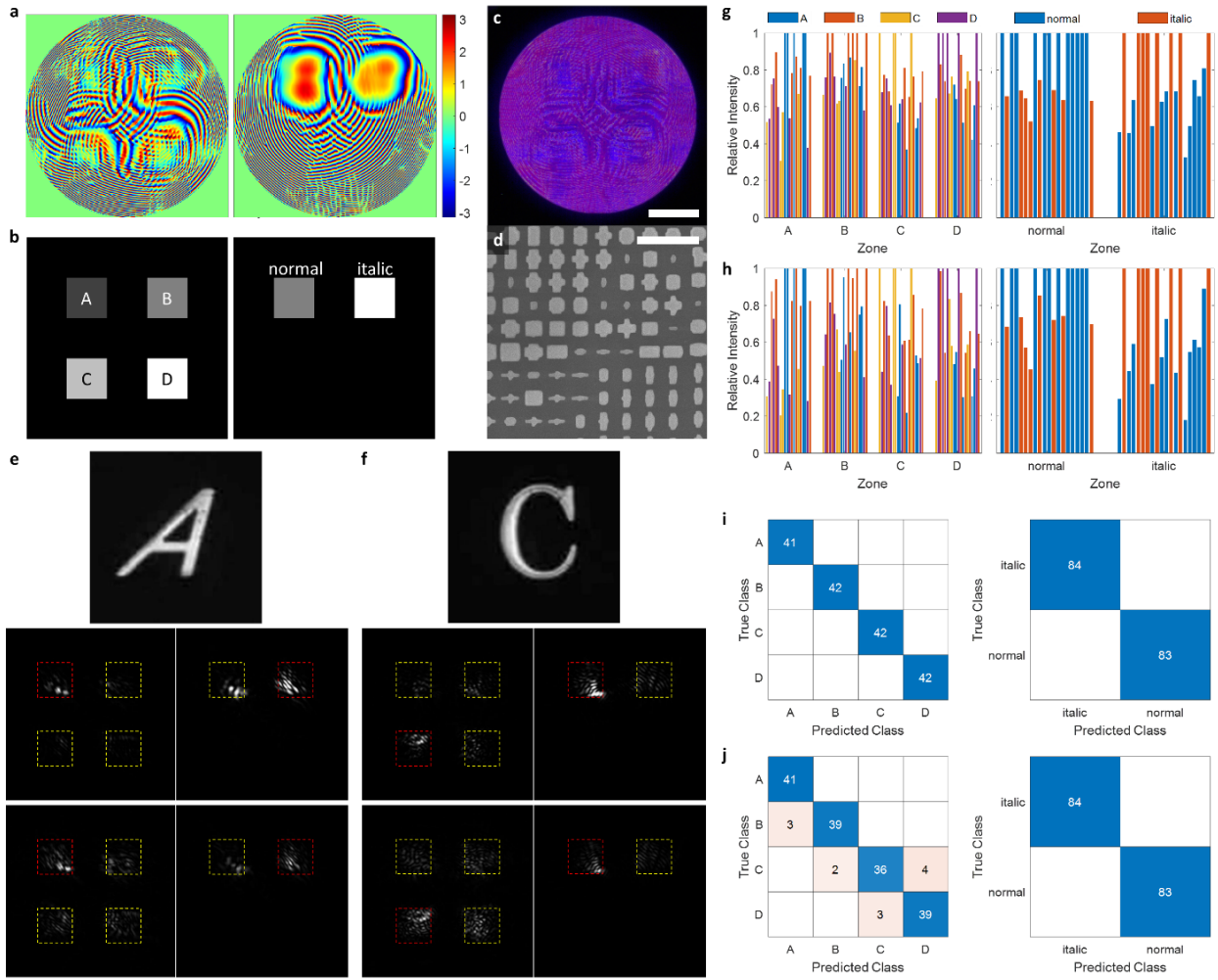

**Figure 5 | Recognition of identity and typographic style of four classes of letters using a polarization-multiplexing smart glass.** (a) Trained phase modulations on the metasurface and (b) arrangements of detection zones on the output plane at two orthogonal incident polarizations for the two distinct recognition tasks. (c) Optical microscopic photo and (d) SEM image of the fabricated metasurface. Scale bars: 200 μm in (c) and 2 μm in (d). (e)-(f) Two examples showing the recognition of an italicized "A" and a normal "C": (Top) Input images; (Middle) Measured intensity distributions on the output plane showing that the detection zone (red) corresponding to the identity (Left) and typographic style (Right) of a letter has the highest integrated optical intensity. (Bottom) Analytically calculated intensity distributions on the output plane showing a high degree of consistency with the measured results. (g) Theoretical integrated intensities of the detection zones for 20 randomly selected input letters, with the highest value normalized to 1. The colors of bars represent the true identities and typographic styles of letters. (h) Experimental

integrated intensities of the detection zones for the same 20 letters as those shown in (g). (i) and (j) Confusion matrices summarizing the theoretical and experimental results of the recognition of 168 letters and their typographic styles, respectively.

**Facial verification using double-layered metasurface smart glass**

Complex recognition tasks beyond digit or letter classification require metasurfaces with enhanced expressive power. To this end, we theoretically demonstrate an ONN consisting of a metasurface doublet for human facial verification (**Fig. 6a**): the ONN can translate an optically coherent, gray-scale image into a low-dimensional representation, allowing one to compare two distinct images of human faces and decide whether the images represent the same person. Specifically, the metasurface doublet maps an image into a 3×3 intensity array on the detection plane, and the similarity between two images is evaluated by calculating the Euclidean distance, or dissimilarity, between the two resulting intensity arrays: if the Euclidean distance is below a threshold, the two images are considered a match (**Fig. 6f**); if the distance is above the threshold, the two images are considered to represent distinct persons (**Fig. 6g**). We use a dataset consisting of photos of 100 people, each person with 14 distinct photos (some examples shown in **Fig. 6b**), for training and testing the ONN: the photos of 90 people are used to train the metasurface doublet and the photos of the remaining 10 people are used in the test. The result shows that when the threshold Euclidean distance is appropriately chosen (e.g., 0.8 in this example), the rate of false acceptance (i.e., percentage of impostor pairs accepted) and the rate of false rejection (i.e., percentage of genuine pairs rejected) are both approximately 10% (**Fig. 6d**), resulting in a verification accuracy of approximately 80%, which is comparable to that achieved by a digital ANN with three convolutional layers (**Fig. 6e**).

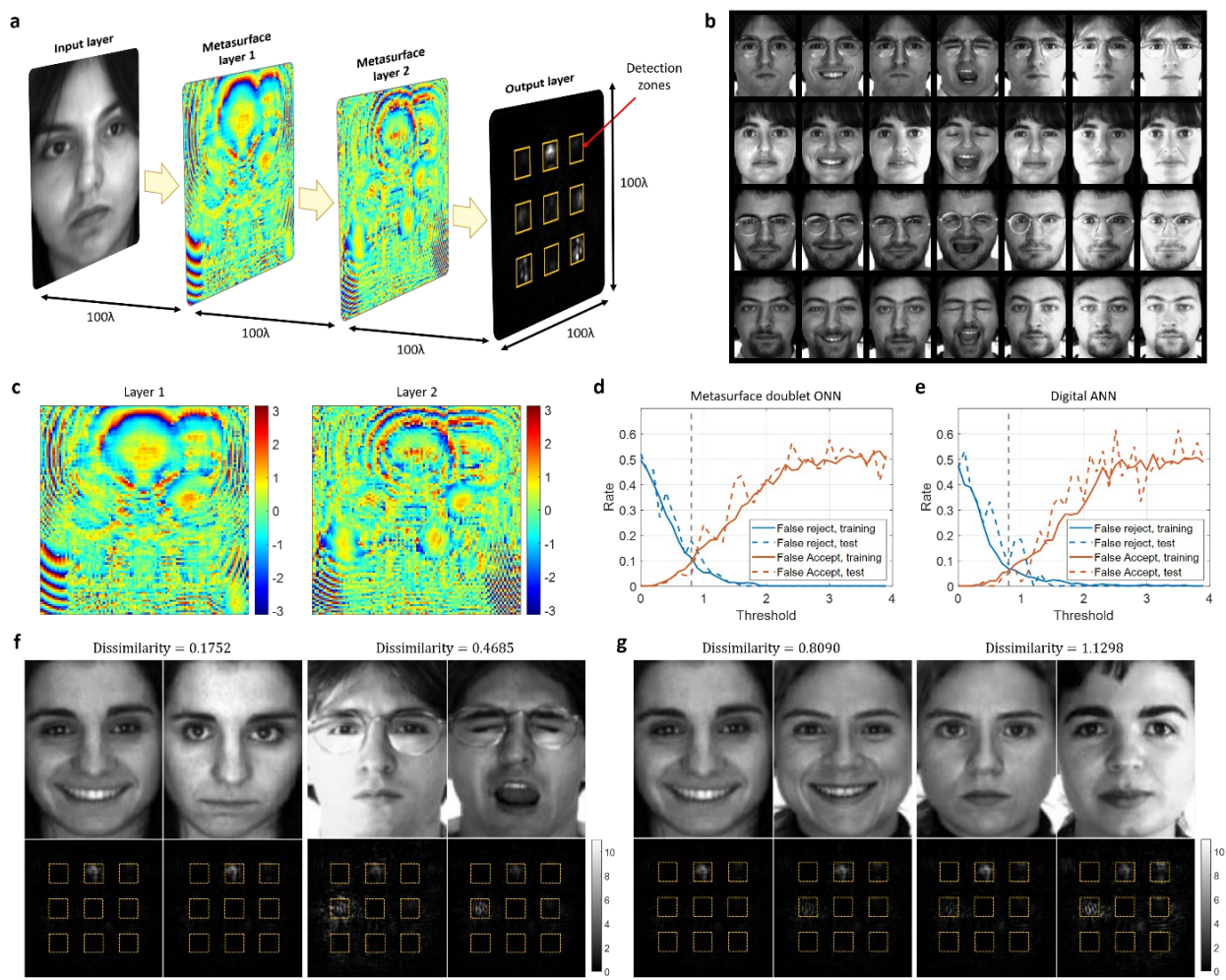

**Fig. 6 | Metasurface doublet for human facial verification.** (a) Schematic illustrating the working mechanism of the metasurface doublet. The metasurface device converts a human face image into a 3×3 optical "barcode", according to the amount of optical power that falls onto the 9 detection zones on the output plane, and whether two images represent the same person is determined by evaluating the difference between their "barcodes". (b) Example human face images used in training and testing the doublet smart glass. (c) Trained phase modulations of the metasurface doublet. (d) and (e) Results showing the evolution of false accept/reject rate (orange/blue curves) as a function of the threshold Euclidean distance of our metasurface doublet ONN and a control digital ANN, respectively. (f) Two examples showing that a pair of photos are determined to represent the same person (despite differences in facial expressions and photo exposure levels) with their Euclidean distance determined by the metasurface doublet below a threshold of ~0.8. (g) Two examples showing that a pair of photos are determined to represent

distinct persons (despite similar facial expressions) because their Euclidean distance is above the threshold.

**Discussion**

**Robustness of metasurface ONN**

We observe that in simulation, an ONN consisting of a single metasurface is usually sufficient to provide a high accuracy of >90% for simple tasks such as digit or letter recognition. However, experiments may report a lower accuracy by a few percent to 20%. This discrepancy is related to the robustness of metasurface smart glasses against experimental errors, and the intensity contrast between the detection zones with the highest and second highest intensities can quantify the robustness of the ONN design. Our experiments show that this inter-zone contrast positively correlates with the degree of agreement between theoretical and experimental recognition accuracies (**Fig. S4**). Thus, by considering this inter-zone contrast in the loss function or by increasing its weight in the loss function while training the ONN, we expect to mitigate the impact of experimental errors on the performance of the ONNs.

**Increasing expressive power of metasurface ONN**

Our experimental results in **Figs. 2-5** indicate that a single metasurface can recognize 4-10 classes of relatively simple optically coherent objects. However, it is a daunting task to create a neural network, even digital ones, to categorize a substantial portion of the images from the ImageNet database [42], which contains over 15 million high-resolution images prelabeled in over 22,000 categories. Our future work will investigate the scalability of the ONN platform to recognize a larger number of classes of objects and to recognize more complex objects. In addition, compared to optically coherent objects, optically incoherent ones are more prevalent in everyday life but represent a bigger challenge for an optical neural network. This is because the cross-product between field components as a result of optical interference provides a form of nonlinear activation in our ONN, and thus, the expressive power of the ONN is reduced in the absence of optical interference when scattered light waves from objects become incoherent.

A general approach to boost the expressive power of the metasurface smart glass is to increase the "width" and "depth" of the ONN. This is a close parallel with the progress in digital ANNs where networks with increased width and depth are developed to solve more complex problems. The ONN depth can be increased by using a multi-layered metasurface architecture; we have demonstrated that the metasurface doublet has enabled recognizing gray-scale images of human faces, which are considerably more complex than binary digits and letters. The ONN width can be increased by employing a few strategies. First, a straightforward method to double the expressive power of a metasurface is to leverage polarization multiplexing. Second, metasurfaces providing complete and independent control of optical phase and amplitude could be more powerful building blocks of an ONN compared to phase-only metasurfaces used in the present work. We have previously demonstrated phase-amplitude metasurface holograms where optical amplitude is controlled by the degree of structural birefringence of meta-units, while optical phase is controlled by the in-plane orientation of the birefringent meta-units [25,32-37]. Another approach to realize simultaneous amplitude and phase control is to use monolithic bilayered meta-units, where silicon and $TiO_2$ can provide amplitude attenuation and phase retardation for visible light, respectively [37]. Third, wavelength-multiplexing can introduce an additional dimension to increase the expressive power of an ONN. The optical dispersion of meta-units (i.e., their phase and amplitude responses as a function of wavelength) can be engineered by controlling the size and shape of the meta-unit cross-sections. As a result, a single metasurface can encode distinct optical amplitude-phase profiles at different wavelengths. Lastly, including an array of distinct metasurfaces in each layer of the neural network is an effective approach to increase its expressive power. Our preliminary investigation indicates that a single layer of 10 distinct metasurfaces is able to classify 10 classes of incoherent objects (i.e., MNIST hand-written digits) with an accuracy higher than 90%.

**Conclusion**

We demonstrated ONNs based on optical metasurfaces that are capable of recognizing binary and gray-scale images with high accuracy. Although our ONNs do not feature a great depth, their expressive power is substantially augmented by the width of each layer due to the millions of subwavelength meta-units in each metasurface. The intrinsic 2D nature and diffraction-based signal processing of the ONN are suitable for applications in object recognition and other image-based computer vision tasks. The width and depth of the ONN can be scaled up to recognize a large number of classes of monochrome and colorful objects illuminated by either coherent or incoherent light. This can be achieved, for example, by using phase-amplitude metasurfaces, by implementing polarization and wavelength multiplexing in each metasurface, by using arrays of metasurfaces on each layer of the network, and by cascading metasurface layers. Aside from leveraging optical interference to introduce a form of nonlinear activation, our ONNs do not utilize the nonlinear activation function in the strict sense as it is implemented in biological and digital neural networks. This fact limits the range of tasks that they can perform and the accuracy that they can achieve. Future work could realize nonlinear activation by introducing nonlinear materials (e.g., semiconductors with saturable absorption) into metasurfaces.

Advanced sensors will be ubiquitous in future applications. These sensors are often deployed in areas or scenarios that lack infrastructure support. They should require minimal service, and feature resilience to interference, high energy efficiency, and information security. These requirements present a daunting challenge for existing technology. An ONN, such as the ones demonstrated in this work, computes directly upon the physical domain, effectively condensing measurement, analog-to-digital conversion, and computing in a single passive device. It uses no power, provides physics-guaranteed security, and has an ultra-compact form factor. Importantly, it can protect the privacy of the subject of interest because there is no representation of the subject in the digital domain. With these advantageous traits, ONNs as "edge" perception devices can fundamentally reshape the future of data collection and analysis.

39. A. Silva, F. Monticone, G. Castaldi, V. Galdi, A. Alu, N. Engheta, "Performing mathematical operations with metamaterials," *Science* **343**, 160−163 (2014).

40. T. Zhu, Y. Zhou, Y. Lou, H. Ye, M. Qiu, Z. Ruan, S. Fan, "Plasmonic computing of spatial differentiation," *Nat. Commun.* **8**, 15391 (2017).

41. A. Cordaro, H. Kwon, D. Sounas, A. F. Koenderink, A. Alù, and A. Polman, "High-Index Dielectric Metasurfaces Performing Mathematical Operations," *Nano Letters* 19, 8418-8423 (2019).

42. ImageNet, https://image-net.org/index.php.